# Study of the solute clusters/enrichment at early stage of ageing in Mg-Zn-Gd alloys by atom probe tomography

Xin-Fu Gu[1,2], Tadashi Furuhara[2], Leng Chen[1], Ping Yang[1]

1. School of Materials Science and Engineering, University of Science and Technology Beijing, Beijing, 100083, China
2. Institute for Materials Research, Tohoku University, Sendai, 980-8577, Japan

*Corresponding author. Xin-Fu Gu, Lecturer, Ph.D.; Tel: +86-10-82376968; E-mail address: xinfugu@ustb.edu.cn

**Abstract**

The chemical enrichment/ordering of solute atoms in Mg matrix are crucial to understand the formation mechanism of long-period stacking ordered (LPSO) structures. In this study, three-dimensional distribution of solute elements in an Mg-Zn-Gd alloy during ageing process is quantitatively characterized by three-dimensional atom probe (3DAP) tomography. Based on the radius distribution function, it is found that Zn-Gd solute pairs in Mg matrix appear mainly in two peaks at early stage of ageing and the separation distance between Zn and Gd atoms is well rationalized by the first-principles calculation. Moreover, the fraction of Zn-Gd solute pairs increases first and decreases subsequently due to the precipitation of LPSO structures. Moreover, the composition of structural unit in LPSO structure and the solute enrichment around it are quantified. It is found that Zn and Gd elements are synchronized in the LPSO structure, and solute segregation of pure Zn or Gd is not observed at the transformation front of the LPSO structure in this alloy. In addition, the crystallography of transformation front is further determined by 3DAP data.

# 1. Introduction

Magnesium alloys are known as the lightest metallic materials for structural application, and thus are attractive to applications in transport system to save the energy consumption by reducing the body weight. However, their applications are limited by their relative low mechanical properties at both room and elevated temperatures and poor corrosion resistance [1]. Recently, a new promising strengthening phase, namely long-period stacking ordered (LPSO) structure, has been found in Mg-Zn-Y and Mg-Zn-Gd alloys [2-8]. This kind of LPSO structure could be also found in Mg-M-RE (M: Zn, Al, Cu, Ni, or Co, RE: Y, Gd, Tb, Dy, Ho, Er, Tm) alloy systems [9-14].

The LPSO structure has a lamellar structure paralleling to $(0001)_\alpha$ basal plane as shown in Fig. 1(a-b), within which four-layer-height fcc structural units (SU for short hereafter, and ABCA stacking ordered SU is highlighted by grey color in Fig. 1) are separated by several Mg layers (hcp structure) depending on the specific structures [14-16]. The hcp structure could be transformed into fcc SU by gliding of a $<10\text{-}10>_\alpha/3$ partial dislocation as schematically shown in Fig. 1(c). The common LPSO structures 10H, 18R, 14H, and 24R correspond to the intermediate Mg layers of 1, 2, 3 and 4 layers, respectively [15, 16], where the symbols for the LPSO structures are Ramsdell's notation. The SUs in LPSO structure usually synchronize with chemical order/enrichment of RE and M [17-19], and it is proposed to contain ordered distributed $TM_6RE_8$ clusters with $L1_2$ type structure [12, 18]. The composition of LPSO structures locates around a line with a M:RE ratio as 3:4 in ternary phase diagram due to the $TM_6RE_8$ clusters in SU, thus theoretical composition for 14H LPSO structure is deduced to be $Mg_{35}TM_3RE_4$ [18, 20], however the measured composition often lower than the theoretical value [12, 18]. The inconsistency could be explained by the existence of domain structures of $L1_2$ clusters in the SU/LPSO [21-23]. Comparing to the crystal structure of LPSO structures, the understanding of transformation mechanism of LPSO structure in crystalline Mg matrix is still unsatisfied [24-27].

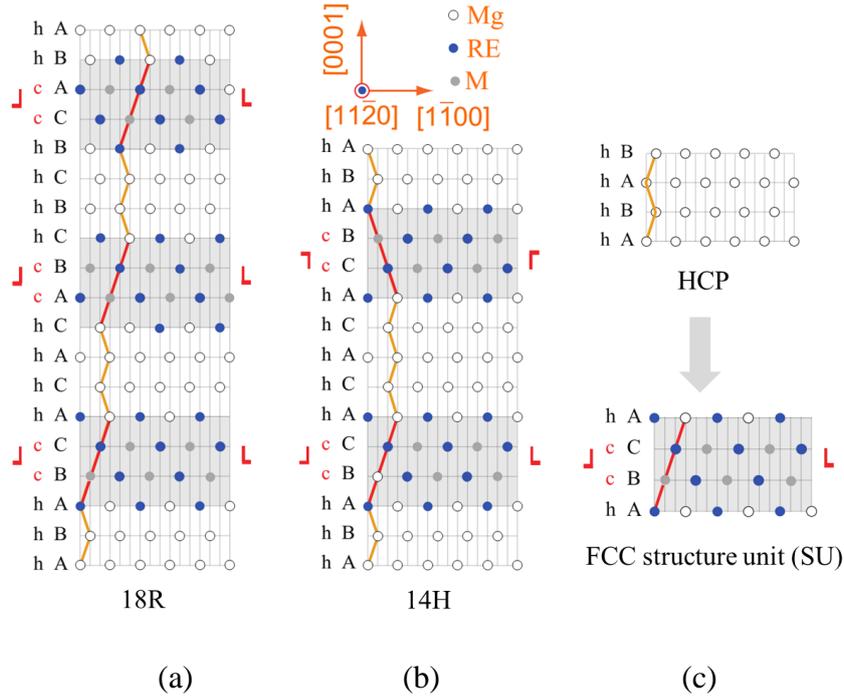

Fig. 1. <11-20>$_\alpha$ view of typical LPSO structures. a) 18R, b) 14H, c) the transformation from hcp structure to fcc SU. The fcc SUs in a-b) is highlighted by grey box. The transformation of hcp structure to fcc needs two processes. One is the stacking sequence change by movement of a partial dislocation, and the other is segregation of RE and M solute atoms to SU.

The LPSO structure in Mg-M-RE alloy is divided into two types depending on the formation conditions [8, 9]. The LPSO phase formed during casting procedure is classified as Type I alloy, while the LPSO phase only generated by subsequent heat treatment after casting is defined as Type II alloy in literature. Likubo *et al.* [28] shows that it probably dues to different spinodal decomposition temperature according to the thermodynamic calculation, and Type II alloy has lower decomposition temperature comparing to Type I alloy. With lower spinodal decomposition temperature, the diffusivity of atoms is not high enough for the transformation, thus LPSO transformation could not be found in the Type II alloy in as-casted condition. Nevertheless, the composition modulation in LPSO phase is thought to be due to the spinodal decomposition during solidification or the subsequent heat treatment.

Meanwhile, theoretically speaking, the transformation sequence of structure change and solute enrichment could vary during formation a fcc SU according to Fig. 1(c), therefore, two possible mechanisms for LPSO formation are possible in crystalline Mg matrix [26, 27, 29], which is schematically shown in Fig. 2. The first mechanism is that the solute atoms often segregate to the dislocations, and this segregation will assist the nucleation of SU, i.e. the structure transformation from hcp to fcc by dislocation dissociation as shown in Fig. 2(a), since the segregation of solute atoms M and RE could lower the stacking fault energy. The second mechanism is that the chemical modulation or cluster takes in first place shown in Fig. 2(b), and then stacking fault forms, and this procedure is repeated for other SUs. Therefore, the cluster behavior in Mg matrix is crucial to distinguish these two processes.

Experimentally, the co-segregation of Zn and Y atoms in Mg-Zn-Y (Type I) alloy, so called Guinier-Preston (G.P.) zone, is observed by high angle annular dark field (HAADF) scanning transmission electron microscopy (STEM) and STEM-EDS [30]. Similar results are found in Mg-Al-Gd (Type I) alloy [25, 29] and Mg-Zn-Gd (Type II) alloy [31]. These results may support the concept of spinodal decomposition, since the observed segregation/enrichment in Mg matrix. On one hand, the observed solute enrichment is often planar type which is observable by HAADF-STEM, on the other hand, three-dimensional distribution of the chemical modulation or solute cluster is not clear. The observation method based on HAADF-STEM cannot quantify the three dimensional chemical information. Moreover, we recently proposed that the concentrated area is mainly Gd according to the experimental result by 3DAP in Mg-Al-Gd (Type I) alloy, i.e. Gd and Al may be not synchronized during LPSO formation [25]. The situation in other alloy system especially for early stage of LPSO formation is unclear. Therefore, the cluster behavior between solute atoms in early stage of ageing should be investigated.

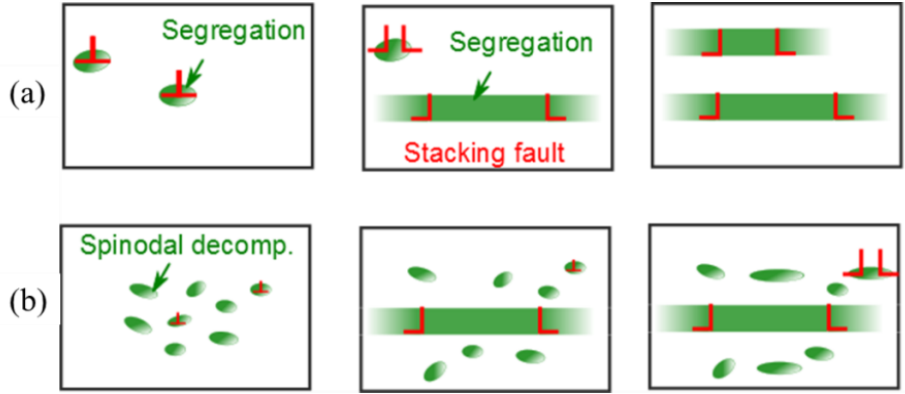

Fig. 2. Schematic diagram of two possible mechanisms for LPSO formation in crystalline Mg matrix. (a) Element segregation assisted dislocation dissociation, (b) Chemical modulation/cluster first without structure change.

Three-dimensional atom probe (3DAP) tomography is a useful method to analyze the spatial distribution of the solute atoms in Mg alloy with relatively good resolution, and it has been used in analyzing the precipitation in Mg alloys, such as precipitates in Mg alloy [32, 33] and clusters in matrix [34, 35]. In this study, the cluster behavior during ageing of $Mg_{97}Zn_1Gd_2$ alloy will be statistacally investigated by 3DAP. Although Nie *et al.* have observed Zn and Gd cluster tendency in a Mg-Zn-Gd alloy, metastable $\gamma''$ phase is formed before the formation of SU ($\gamma'$) [34]. On contrary, in present study with different alloy composition, SU directly precipitates from the matrix without precursor phase.

**2. EXPERIMENTAL PROCEDURE**

Since $Mg_{97}Zn_1Gd_2$ alloy is a Type II alloy, there are little LPSO pre-exist in the as-casted sample, moreover ageing of the sample after solution treatment would precipitate precursor $\gamma''$ phase other than LPSO SU at such temperature [34, 36-38], thus direct ageing of cast sample is carried out at 280 ℃ for 15min and 1h. In 3DAP technique, the atoms in the needle-like sample are evaporated one by one by the evaporation field due to curvature. The spatial resolution is anisotropic in 3DAP, and the resolution along the needle axis is the best, which is smaller than 0.25nm (spacing between basal plane). Therefore, in order to observe the chemical distribution for LPSO structure, the needle axis should be closely paralleling to $(0001)_\alpha$ plane to reach highest resolution. Therefore, we find proper grains with the aid of electron backscatter diffraction (EBSD) system (TSL, EDAX). The procedure for preparing a 3DAP sample within Mg matrix is shown in Fig. 3(a-f) in order. Firstly, the grain with normal direction (ND) close to $(0001)_\alpha$ is selected by EBSD mapping. Then the needle sample is cut and needled by focus ion beam (FIB) system equipped in scanning electron microscope (SEM, Quanta 200, FEI). The 3DAP data is taken with CAMECA Leap 4000HR at 35K. A pulse fraction of 20% and a pulse frequency of 200 kHz are used. The cluster tendency is analyzed by radius distribution function.

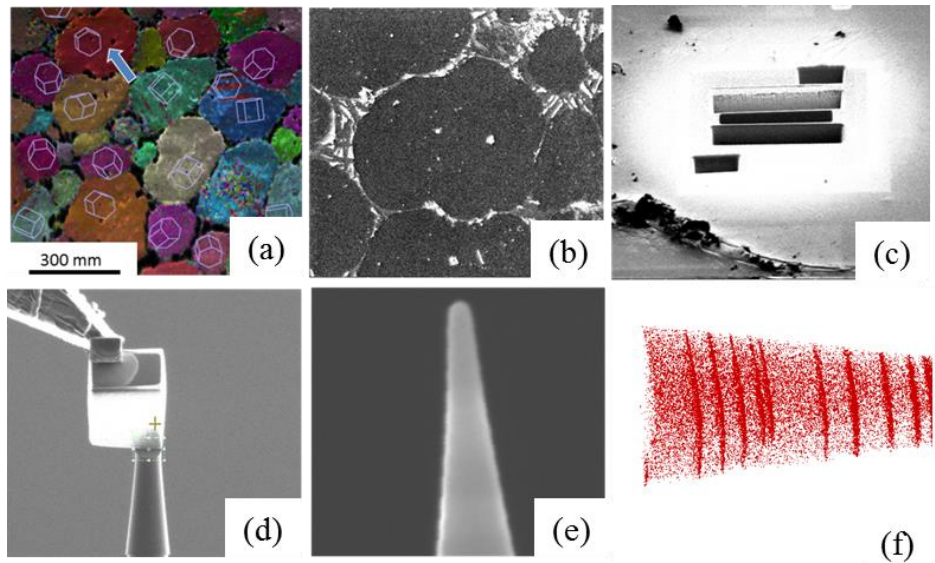

Fig. 3. 3DAP sample preparation procedure (a) Inverse pole figure (Z) mapping of Mg sample by EBSD, (b) Selected grain as indicated in (a) with arrow, (c) Cutting with FIB, (d) Mounting to silicon stage, (e) Needling, (f) Test and obtain the result of element distribution.

Table 1 Composition in Mg matrix before and after solution treatment

| Atom (at.%) | Mg | Gd | Zn |
|---|---|---|---|
| Un-solution | 99.4% | 0.40% | 0.24% |
| Solution | 98.3% | 1.3% | 0.39% |

## 3. RESULTS AND DISCUSSION

Fig. 4(a-b) shows the 3DAP result of Mg-Zn-Gd for as-casted sample and aged sample for 1h at 280 °C respectively. Only the element mapping of Zn and Gd atoms are shown here, where Gd atoms are shown by green color while Zn atoms are shown by blue colors. In Fig. 4(a), there are no precipitates found in the Mg matrix in the as-casted sample. Table 1 shows the composition in Mg matrix in as-casted state. Comparing to solution treated sample at 520°C for 2h, Zn and Gd contents in as-casted sample is more dilute, and Gd/Zn ratio is also low. This composition difference may cause different precipitation before and after solution treatment. The precipitation in solution treated sample is mainly γ″, while the precipitates in as-casted sample is SU (γ′) [31]. Therefore, in order to directly study Zn-Gd clusters before the precipitation of SU, direct ageing of as-casted sample is investigated in this study.

In Fig. 4(b), obvious enrichment of solute atoms is found after ageing. According to 3-D distribution of solute atoms, the shape of the enrichment in (b) is confirmed to be planar shape. These kind of enrichment is due to the precipitation of fcc SU when ageing 1h at 280 °C [31]. In the SU, they are synchronized with chemical enrichment of Zn and Gd according to Fig. 4(b). In addition, the SU is parallel to $(0001)_\alpha$ plane, and the needle axis is nearly along $[0001]_\alpha$.

A detailed view of rectangular area in Fig. 4(b) is shown in Fig. 5. The pink dots represent Mg atoms. The spacing of Mg layers in Fig. 5(a) is about 0.5 nm, which is near the lattice parameter $c$ of hcp Mg. In the left side of Fig. 5(a), there are two SUs with spacing of about 2 nm which corresponding to 14H LPSO. The right side of Fig. 5(a) shows a single SU. From these single SUs, the composition of SU could be determined. Fig. 5(b) shows the concentration profile of Zn and Gd in Fig. 5(a). Zn and Gd atoms are synchronized in the SUs. It is also found that the composition of SU is $MgGd_{4\sim7}Zn_{3\sim6}$, which is lower than ideal composition of SU, i.e. $Mg_{70.8}RE_{16.7}Zn_{12.5}$ deduced from perfect LPSO structure from Egusa *et al* [5]. Based on the compositions, it could be further confirmed the precipitates in present study are SUs here, since the contents of Zn and Gd of γ″ phase is near 15 at. % while SU has lower Zn and Gd content [34, 39]. Considering of the composition variations in SUs in Fig. 5(b), it could be deduced that the solute atoms in SU may be enriched further during ageing. Taking the left two neighboring SUs in Fig. 5(b) for example, the size of SU in left side is smaller than right one according to Fig. 4(b), and probably the left SU is formed late. Correspondingly, the composition of Zn and Gd in left SU is lower than right one.

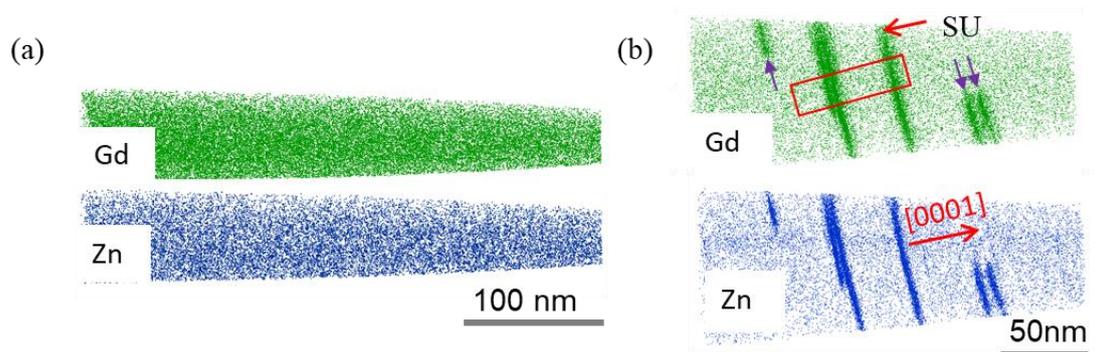

Fig. 4. Element distribution of Gd and Zn solute atoms in Mg matrix for Mg-Zn-Gd alloy. (a) As-casted sample, (b) Aged for 1h at 280 °C.

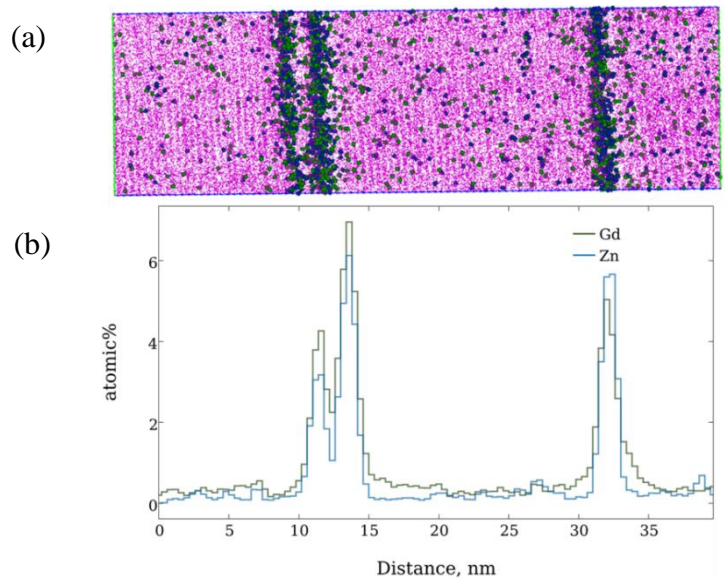

Fig. 5. Enlarged view of rectangular area in Fig. 4(b). (a) Element mapping, Mg atoms are shown by pink dots, (b) Corresponding concentration profile.

Since the element distributions at different ageing time have been obtained by 3DAP tomography, the solute clusters within the matrix could be further analyzed. Fig. 6 shows the evolution of the clusters between Zn and Gd during ageing. The cluster tendency between Zn and Gd in matrix is evaluated by the radial distribution function. For calculating the radial distribution function, it calculates the number of Gd atoms in a shell of width $\Delta r = 1$ Å at distance $r$ with a given atom Zn, and the number is normalized with average density. For random distribution, the density should be 1, and any deviation from 1 will indication a cluster or separation tendency. This method is effective for small clusters [40]. According to Fig. 6, it is found that Zn and Gd atoms have the tendency to stay together to be a cluster, because the normalized density is higher at smaller separation distance. At the early stage of ageing, the shortest clustering distance between Zn and Gd atoms is about 0.2~0.4 nm according the radial distribution between those atoms, which is close to the lattice parameter of Mg matrix in base plane (0.32 nm). Furthermore, a high frequency at the distance of around 0.8~1 nm is recognized, which may indicate some correlation between clusters. As ageing is prolonged, the SUs form, and the cluster tendency between Zn and Gd atoms is weakened, which may due to the precipitation of SU and LPSO structure.

According to the first-principle calculation by Kimizuka *et al* [41], the binding energy of different atomic pairs in matrix varies at different nearest neighbor (NN) as shown in Table 2. The different NN is schematically shown in Fig. 7. For Zn-Gd pair, they are preferred at short NN, and may correspond to the 0.2-0.4 nm peak in Fig. 6. The near 1.0 nm peak may correspond to the distance between different $L1_2$ clusters after formation of the SU [42]. Therefore, the solute atoms have the tendency to form Zn-Gd clusters in the Mg matrix before the formation of SU. Next, the solute enrichment near the interface between SU and Mg matrix will be studied.

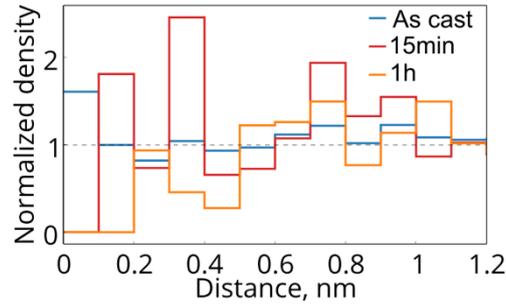

Fig. 6. Radius distribution function between Zn and Gd pairs in Mg matrix for different ageing time.

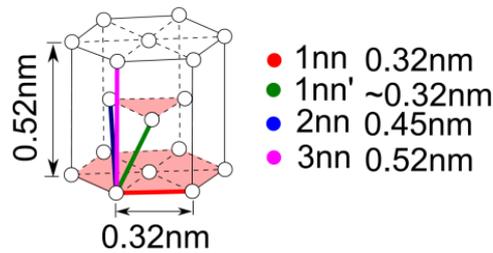

Fig. 7. The interatomic distance for different nearest neighbors in hcp structure.

Table 2 The binding energy between solute atoms in the matrix [41]

|      | Zn-Zn  | Gd-Gd  | Zn-Gd  |
|------|--------|--------|--------|
| 1nn  | 0.004  | 0.022  | -0.077 |
| 1nn' | 0.011  | 0.066  | -0.08  |
| 2nn  | -0.026 | -0.172 | 0.064  |
| 3nn  | 0.0001 | -0.104 | 0.015  |

Transformation (growth) fronts in the SUs are observed in both Fig. 4(b), and they are indicated by double arrows in the figure. A thin slice parallel to $(0001)_\alpha$ around the left SU in Fig. 4(b) is taken out and viewed along $[0001]_\alpha$ direction as shown in Fig. 8. The transforming SU is clear shown in Fig. 8(b-c), since it is enriched with Gd and Zn. The $[0001]_\alpha$ zone in Fig. 8(a) could be found in Mg atom map according to its symmetry, and this pattern could be used to identify the edge direction of the transformation front. Since the needle axis of 3DAP sample is close to $[0001]_\alpha$, the pole of $[0001]_\alpha$ can be easily found at the detector event histogram which is due to the aberration evaporation near zone axis [43, 44]. The superimposition of detector event histogram and stereographic projection is shown in Fig. 9, and the aberration traces are along $<1\text{-}100>_\alpha$ direction. Therefore, the traces in Fig. 8(a) are along $<1\text{-}100>_\alpha$ direction according to Fig. 9. The $<1\text{-}100>_\alpha$ direction is frequently observed in our case, though the edge of transformation front along $<11\text{-}20>_\alpha$ direction is also observed, which is consistent with TEM study [45]. The variation of composition normal to the edge of transformation front, i.e. along the arrow direction in Fig. 10 (b), is shown in Fig. 10(a). Gd and Zn are synchronized along the direction indicated by the arrows. The component of Zn and Gd is similar to the left SU in Fig. 5, which means the initial formed SU may have lower solute content and will be enriched after further annealing. Nevertheless, the growth of LPSO is accompanied by the solute enrichment of both Zn and Gd solute atoms.

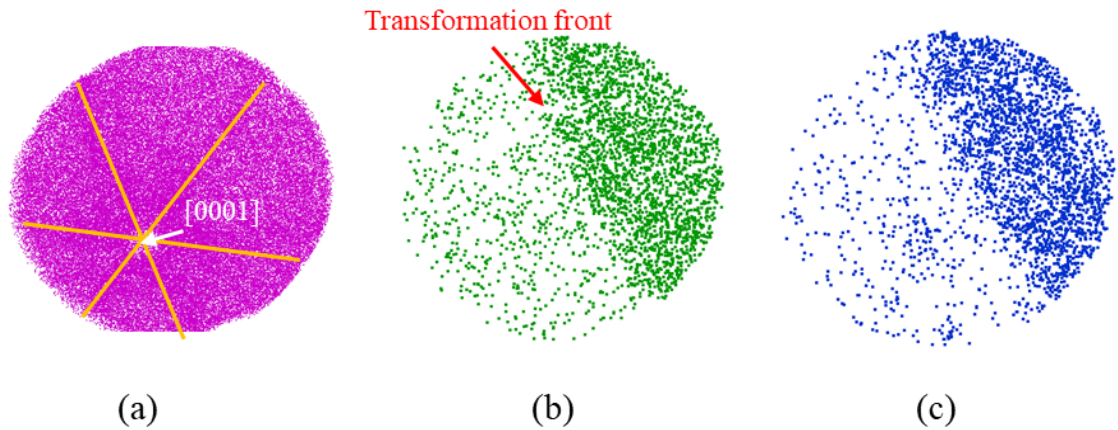

Fig. 8. The transforming front of the SU in the left side in Fig. 4(b). (a) Mg mapping, (b) Gd mapping, (c) Zn mapping, (d) element distribution along the arrow in (b). The view direction is $[0001]_\alpha$.

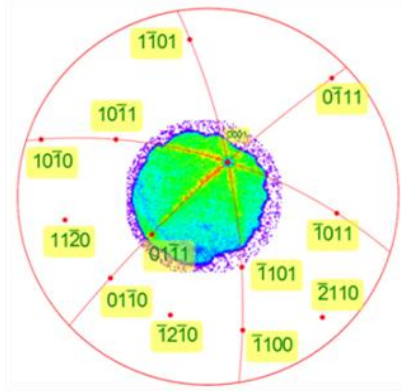

Fig. 9. Superimposition of detector event histogram and stereographic projection of hcp structure.

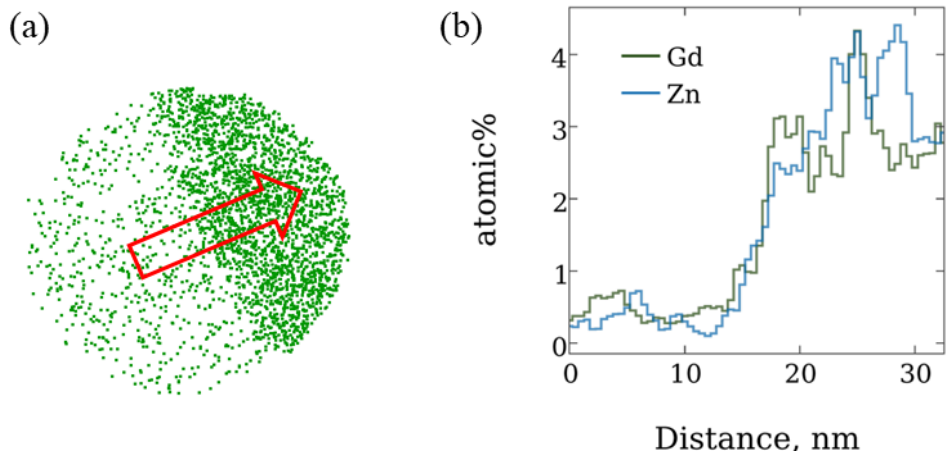

Fig. 10. The composition across the transformation front. (a) Element distribution of Gd, (b) Concentration profile.

## 4. CONCLUSIONS

The chemical enrichment of solute atoms in Mg matrix and LPSO structure in $Mg_{97}Zn_1Gd_2$ has been studied by 3DAP for different ageing time.

(1) According to the radial distribution between Zn and Gd in Mg matrix, solute atoms Zn and Gd tend to be cluster at first nearest neighbor and decreases with ageing time due to precipitation. The cluster behavior could be rationalized by the binding energy between Zn and Gd atoms.

(2) The fcc SU are synchronized with chemical enrichment of Zn and Gd. The composition of SU is determined to be $MgGd_{4\sim7}Zn_{3\sim6}$ and lower than theoretical composition. The content of Zn and Gd in SU enriches during ageing.

(3) The complementary crystallography information on transformation fronts have been obtained from 3DAP data. The edge of transformation front in a SU is often along low indexed direction, such as $<1\text{-}100>_\alpha$, according to the aberration evaporation near $[0001]_\alpha$ zone axis. The solute atoms Zn and Gd are also synchronized at the front of the transformation in this system, and the growth of LPSO structure is accompanied by the solute enrichment of both Zn and Gd solute atoms.

**ACKNOWLEDGEMENTS**

This work was supported by a Grant-in-Aid for Scientific Research on Innovative Areas, "Synchronized Long-Period Stacking Ordered Structure", from the Ministry of Education, Culture, Sports, Science and Technology, Japan (No.23109006) and Fundamental Research Funds for the Central Universities (No. FRF-TP-17-003A1). Last but not least, the invitation of Prof. R.-Z. Wu (Harbin Engineering University) to this symposium is greatly appreciated.

Tables

Table 1. Composition in matrix before and after solution treatment.

Table 2. The binding energy between solute atoms in the matrix [41]

Figures

Fig. 1. $\langle 11\text{-}20\rangle_\alpha$ view of typical LPSO structures. a) 18R, b) 14H, c) the transformation of hcp structure to fcc SU. The fcc SUs in a-b) is highlighted by grey box. The transformation of hcp structure to fcc needs two processes. One is the stacking sequence change by movement of a partial dislocation, and the other is segregation of RE and M solute atoms to SU.

Fig. 2. Schematic diagram of two possible mechanisms for LPSO formation in crystalline Mg matrix. (a) Element segregation assisted dislocation dissociation, (b) Chemical modulation/cluster first without structure change.

Fig. 3. 3DAP sample preparation procedure (a) Inverse pole figure (Z) mapping of Mg sample by EBSD, (b) Selected grain as indicated in (a) with arrow, (c) Cutting with FIB, (d) Mounting to silicon stage, (e) Needling, (f) Test and obtain the result of element distribution.

Fig. 4. Element distribution of Gd and Zn solute atoms in Mg matrix for Mg-Zn-Gd alloy. (a) As-casted sample, (b) Aged for 1h at 280 °C.

Fig. 5. Enlarged view of rectangular area in Fig. 4(b). (a) Element mapping, Mg atoms are shown by pink dots, (b) Corresponding concentration profile.

Fig. 6. Radius distribution function between Zn and Gd pairs in Mg matrix for different ageing time.

Fig. 7. The interatomic distance for different nearest neighbors in hcp structure.

Fig. 8. The transforming front of the SU in the left side in Fig. 4(b). (a) Mg mapping, (b) Gd mapping, (c) Zn mapping, (d) element distribution along the arrow in (b). The view direction is $[0001]_\alpha$.

Fig. 9. Superimposition of detector event histogram and stereographic projection of hcp structure.

Fig. 10. The composition across the transformation front. (a) Element distribution of Gd, (b) Concentration profile.